\documentclass{iaus}
\usepackage{graphicx}
\usepackage{natbib}
\usepackage{amsmath}
\usepackage[font=scriptsize]{subfig}

\title[An extra-galactic planet around an old red-horizontal branch star]{The visitor from an ancient galaxy: A planetary companion around an old, metal-poor red horizontal branch star}

\author[Rainer J. Klement, Johny Setiawan et al.]   
{Rainer J. Klement$^1$, Johny Setiawan$^1$, Thomas Henning$^1$, Hans-Walter Rix$^1$, Boyke Rochau$^1$, Jens Rodmann$^2$, Tim Schulze-Hartung$^1$}

\affiliation{$^1$Max-Planck-Institute for Astronomy\\ K{\"o}nigstuhl 17,
D-69121 Heidelberg, Germany \\ email: {\tt klement@mpia.de}\\
$^2$European Space Agency, Space Environment and Effects Section, ESTEC} 

\pubyear{2010}
\volume{276}  
\pagerange{}
\jname{The Astrophysics of Planetary Systems: Formation, Structure, and Dynamical Evolution}
\editors{A. Sozzetti, M.G. Lattanzi \& A.P. Boss, eds.}
\begin{document}

\maketitle

\begin{abstract}
We report the detection of a planetary companion around \mbox{HIP 13044}, 
a metal-poor red horizontal branch star belonging to a stellar halo stream that results from the disruption of an ancient Milky Way satellite galaxy. 
The detection is based on radial velocity observations with FEROS at the 2.2-m MPG/ESO telescope. The periodic radial velocity 
variation of $P=16.2$ days can be distinguished
from the periods of the stellar activity indicators.
We computed a minimum planetary 
mass of 1.25 M$_\mathrm{jup}$ and an orbital semimajor axis 
of 0.116 AU for the planet.
This discovery is unique in three aspects: 
First, it is the first planet detection around a star with a metallicity 
much lower than few percent of the solar value; second, 
the planet host star resides in a stellar evolutionary stage 
that is still unexplored in the exoplanet surveys; 
third, the planetary system \mbox{HIP 13044} most likely has an extragalactic origin in 
a disrupted former satellite of the Milky Way.
\end{abstract}

\firstsection 
\section{Introduction}
Do planets exist in external galaxies? There is no reason that they should not. However, directly observing stars in one of the Milky Way's satellites in order to obtain precise enough radial velocity or transit photometry measurements to detect planetary signals is not possible because the galaxies are too far away.\footnote{With a distance modulus of $\sim$14.5, the Canis Major dwarf galaxy could pose an exception, allowing at least principally for planet searches around giant stars; but it is located behind the Milky Way's disk ($l=240^\circ$, $b=-8^\circ$) and therefore heavily obscured by dust and foreground light \citep{mar04,bel04}.} Although microlensing would in principal allow for extra-galactic planet detections, this phenomenon is intrinsically non-repeatable, and no microlensing planets in other galaxies have been confirmed yet \citep[see, however,][]{ing09}. Nevertheless, we know that certain stars in our solar neighborhood have been accreted from satallite galaxies in the past and in fact, examples of such accretion events are still observable today \citep[e.g.][]{bel06}. Accreted stars therefore provide an opportunity to indirectly search for planets that originated in external galaxies and travelled into our Galaxy along with their host stars. In particular, stars belonging to nearby stellar halo streams are ideal targets to search for extra-galactic planets. In the inner halo, where the orbital frequencies are high, such streams are no longer spatially coherent, but they still share similar chemical and dynamical properties that depend on the composition and orbit of their progenitor system and allow for identifying individual stream members \citep[see e.g. the review by][]{kle10}.

The most significant stellar halo stream in the solar neighborhood is the one discovered by \citet{hel99}. Its members have kinematics that clearly separate them from the bulk of other halo stars. The stream's progenitor system possibly resembled a galaxy similar to Fornax or the Sagittarius dwarf, and has been disrupted about 6--9 Gyr ago \citep{hel99,kep07}.
The Red Horizontal Branch (RHB) star \mbox{HIP 13044} is a confirmed member of the Helmi stream \citep{hel99,ref05,kep07,roe10}.
Its stellar parameters, in particular the low metallicity and relatively large projected rotational velocity (Table\,\ref{tab:t1}), mark this star as an interesting target for a planet search.

\begin{table}
  \begin{center}
  \caption{Important stellar parameters of \mbox{HIP 13044}.}
  \label{tab:t1}
 {
  \begin{tabular}{llll}
\hline
\hline  
  Parameter      &  Value		&  Unit   	& Reference \\
 \hline
 Spectral type      &  F2		&     		& {\sc SIMBAD}\\
 $m_{V}$            &  9.94 		& mag 		& {\it Hipparcos}\\
 $T_{\mathrm{eff}}$ &  6025$\pm$63	& K     	& \citet{car08a,roe10}\\
 $R_*$		    &  6.7$\pm$0.3 	& R$_{\odot}$ 	& \citet{car08a}\\
 {[Fe/H]}     &  $-2.09\pm0.26$	&  & \citet{bee90,chi00};\\
              &              &  &  \citet{car08b,roe10}\\
 $v \sin{i}$	    &  8.8$\pm$0.8   	& $\mathrm{km\,s}^{-1}$ & \citet{car08a} \\
 		    &  11.7$\pm$1.0   	& $\mathrm{km\,s}^{-1}$ & own measurement\\
 \hline 
 \hline
 \end{tabular}
  }
 \end{center}
 \end{table}

\section{Data and Analysis}
We have obtained 36 radial velocity (RV) measurements of \mbox{HIP 13044} between September 2009 and July 2010 with FEROS,
a high-resolution spectrograph ($R$ = 48,000) attached at the 2.2 meter
Max-Planck Gesellschaft/European Southern Observatory telescope, 
located at the La Silla observatory in Chile \citep{kau00}.
RV values have been obtained through cross-correlating the
stellar spectrum with a numerical template designed for stars of the spectral type F0 and containing 550 selected spectral lines.
Typical uncertainties of the RV values are $\sim50$ m s$^{-1}$.
 
\begin{figure}[b]
\begin{center}
\subfloat[]{
 \includegraphics[width=0.45\textwidth]{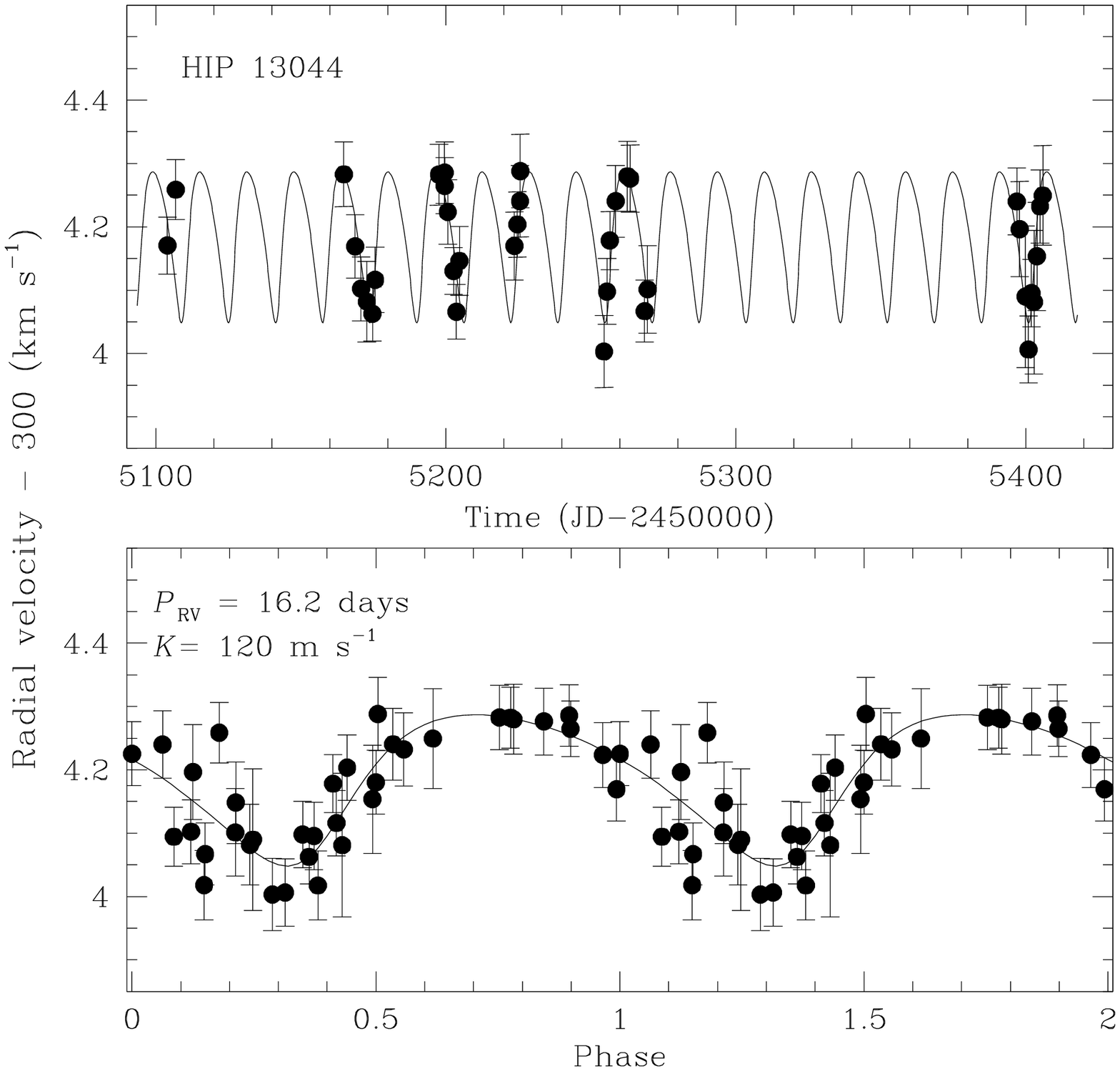}\label{fig:f1a}}
 \quad
 \subfloat[]{
 \includegraphics[width=0.45\textwidth]{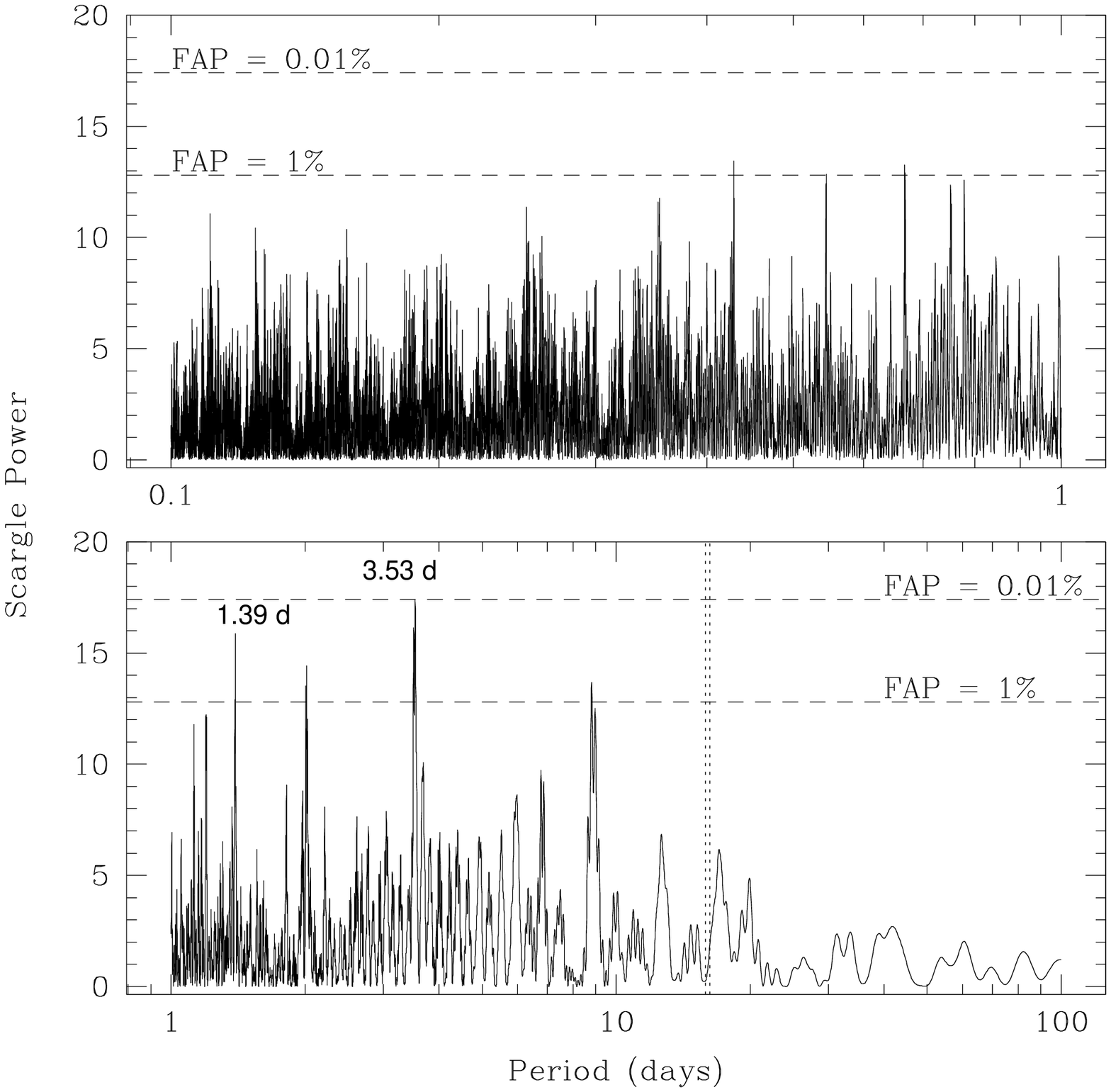}\label{fig:f1b}}
 \caption{(a) RV variation of \mbox{HIP 13044}. (b) LS-periodogram of the \textit{Super-WASP} data.}
   \label{fig:f1}
\end{center}
\end{figure}

The variation of the RV between our observations at different epochs has a 
semi-amplitude of 120 $\mathrm{m\,s}^{-1}$ (Fig.\,\ref{fig:f1a}).
The Generalized Lomb Scargle (GLS) periodogram \citep{zec09}
reveals a significant RV periodicity at $P=16.2$ days with 
a False Alarm Probability (FAP) of $5.5\times10^{-6}$. 
Additional analysis by a Bayesian algorithm \citep{gre05} confirms this period. 
To find out whether stellar activity (moving/rotating surface inhomogeneities or
stellar pulsations) is responsible for producing the observed RV variation, we investigated different stellar activity indicators.

We analyzed the variation of two independent spectroscopic activity indicators: the bisector velocity spans \citep[][]{hat96} of all 550 lines and
the equivalent width (EW) of the infrared CaII line at $\lambda=849.8$ nm.
The GLS periodogram analyses for both the BVS and the Ca II line EW variations revealed significant peaks at similar periodicities ($P_\text{BVS}=5.02$ d, $\text{FAP}_\text{BVS}=1.4\times10^{-5}$ and $P_\text{EW}=6.31$ d, $\text{FAP}_\text{EW}=4.4\times10^{-6}$). We attribute these variations in the spectral line indicators
to rotational modulation caused by star spots or large granulation cells. We point out that we have found no hints for a period around 16 days in the BSV and EW variations.  
  
Photometric observations of \mbox{HIP 13044} have been made by \textit{Super-WASP}. There exist 3620 high-precision measurements of \mbox{HIP 13044} in the public \textit{Super-WASP} archive. After removing 10 data points with high error bars, the Lomb-Scargle periodogram shows that there is no signal for a $\sim$ 16 d period (Fig.\,\ref{fig:f1b}). Instead, there are several marginally significant peaks between a few hours and a few days ($\text{FAP}<1$\%), with the ones at 1.4 d and 3.5 d beeing the most significant. These two peaks are, however, most likely harmonic to each other: $1.4^{-1}+3.5^{-1}=1$. It is expected that \mbox{HIP 13044} oscillates only at pulsationally unstable overtones of high order \citep{xio98}. Observations of one RHB star in the metal-poor globular cluster NGC 6397 \citep{ste09} as well as theoretical predictions \citep{xio98} set these periods in the range of a few hours to a day or so. We caution, however, that no clear theoretical predictions for a star with parameters similar to \mbox{HIP 13044} exist, and it could be possible that some high-order oscillations are able to explain the 1.4 or 3.5 day signal. More important, however, is that we found no signal of a period around 16.2 d in the photometric data.

\section{Discussion and Conclusions}

From the lack of a $\sim16$ d period in both photometric and spectroscopic activity indicators, we conclude that a (sub-stellar) companion remains the most likely hypothesis for the observed $16.2$ d RV variation. Table\,\ref{tab:t2} lists the orbital solution for \mbox{HIP 13044 b} that we have computed from the stellar reflex motion. While the semi-major axis is not unusual for a giant extra-solar planet, the non-circular orbit ($e=0.25$) for such a close-in companion is. One has to keep in mind, however, that the star's red giant branch (RGB) phase, which preceeds the RHB phase, probably has changed the orbital properties of the planet.
From the lack of close-in \citep[$\lesssim0.5$ AU, e.g.][]{sat08} giant planets around RGB stars, it seems plausible that such planets, if they existed, have been engulfed during the expansion of the stellar atmosphere \citep[although other explanations exist, e.g.][]{cur09}. Such planet-swallowing would spin up the star and increase the mass loss \citep{sok98,car09}, and indeed enhanced rotation among RGB and RHB stars has been found \citep{car03,car08a}. Interestingly, \mbox{HIP 13044} is such a fast rotator \footnote{$v_\text{rot}\approx60$ km s$^{-1}$, following from $P_\text{rot}/\sin i=2\pi R_*/(v_\text{rot}\sin i)$, with $R_*$ from Table\,\ref{tab:t1} and $P_\text{rot}\approx5.5$ d from the BVS and EW variations.} too, so possibly \mbox{HIP 13044 b} was part of a multiple planetary system of which some planets have not survived the RGB phase. The survival of \mbox{HIP 13044 b} during that phase is theoretically possible under certain circumstances \citep{sok98}. It is also possible that the planet's orbit decayed through tidal interaction with the stellar envelope. However, a prerequisite to survival is then that the mass loss of the star stops before the planet would have been evaporated or accreted. Assuming asymmetric mass loss, velocity kicks could have increased the eccentricity of \mbox{HIP 13044 b} to its current, somewhat high value \citep{hey07}. The same could be achieved by interaction with a third body in the system. The formation of the planet(s) around such a metal-poor star is possible via gravitational instability. Even if we assume that the star's metallicty was higher in the past, the fact that it belongs to the chemically homogenous Helmi stream \citep[$\langle\text{[Fe/H]}\rangle=-1.8\pm0.4$ for 33 stream members from][]{kep07,kle09} still implies the lowest metallicty of a planet host star found to date. The reason why most planet searches around metal-poor stars failed could be twofold: first, most planets around such stars should reside at large orbital distances \citep[e.g.][]{mar07,san10}; we could have been lucky that the RGB phase lead to an orbital decay of \mbox{HIP 13044 b}. Second, most planet searches tried to observed a large sample simultaneously \citep[e.g.][]{soz09}, in this way eventually not obtaining enough data points for a significant RV signal.

We can be pretty sure that the planet was not captured from another star inside the Milky Way, because the time on which stellar encounters play a role exceeds the Hubble time for galaxies like the Milky Way \citep{bin08}. The stream membership of \mbox{HIP 13044} therefore implies an extra-Galactic origin of \mbox{HIP 13044 b}. 

\begin{table}[ht]
 \begin{center}
 \caption{Orbital parameters of \mbox{HIP 13044 b} }
  \label{tab:t2}
   \begin{tabular}{lll}
 \hline
 \hline 
 $P$      	    &  16.2    $\pm$ 0.3 	    & days  \\
 $K_{1}$            &  119.9   $\pm$ 9.8	    & $\mathrm{m\,s}^{-1}$\\
 $e$                &  0.25    $\pm$ 0.05	    &	    \\
 $\omega$           &  219.8   $\pm$ 1.8	    & deg   \\
 $JD_{0}-2450000$   &  5109.78 $\pm$ 0.02	    & days  \\
 $\chi^{2}$ 	    &  32.35 		    	    & $\mathrm{m\,s}^{-1}$\\
 rms        	    &  50.86     		    & $\mathrm{m\,s}^{-1}$\\
 $m_{2} \sin{i}$     &  1.25    $\pm$ 0.05	   & M$_{\mathrm{Jup}}$ \\
 $a$	            &  0.116   $\pm$ 0.01	   & AU    \\
 \hline 
 \hline
 
 \end{tabular}
 \end{center}
\end{table}

\end{document}